\documentclass[reprint,amsmath,superscriptaddress,amssymb,aps]{revtex4-1}
\usepackage{graphicx}
\usepackage{dcolumn}
\usepackage{bm}
\usepackage{hyperref}
\usepackage{color}

\DeclareMathOperator{\sign}{sign}

\begin{document}

\preprint{APS/123-QED}

\title{Temporal control of graphene plasmons}


\author{Josh Wilson}
\affiliation{School of Mathematics, University of Minnesota, Minneapolis, MN 55455, USA}
\author{Fadil Santosa}
\email{santosa@umn.edu}
\affiliation{School of Mathematics, University of Minnesota, Minneapolis, MN 55455, USA}
\email{santosa@umn.edu}
\author{Misun Min}
\affiliation{Mathematics and Computer Science Division, Argonne National Laboratory, Argonne, IL 60439, USA}
\author{Tony Low}
\affiliation{Department of Elec. \& Comp. Engineering, University of Minnesota, Minneapolis, MN 55455, USA}
\date{\today}

\begin{abstract}
Electrostatic gating and optical pumping schemes enable efficient time modulation of graphene's free carrier density, or Drude weight. We develop a theory for plasmon propagation in graphene under temporal modulation. When the modulation is on the timescale of the plasmonic period, we show that it is possible to create a backwards-propagating or standing plasmon wave and to amplify plasmons. The theoretical models show very good agreement with direct Maxwell simulations.  
\end{abstract}

\pacs{Valid PACS appear here}
                             
\maketitle

\emph{\label{sec:intro}Introduction} -- Two-dimensional layered materials have been intensively explored in recent years for their enhanced light-matter interactions through a plethora of dipole-type excitations\cite{avouris20172d, low2017polaritons, basov2016polaritons}. Graphene, in particular, can accomodate electrically tunable and highly confined low loss plasmon-polaritons\cite{chen2012optical, fei2012gate, koppens2011graphene, low2014graphene, garcia2014graphene, low2017polaritons, basov2016polaritons}. The plasmonic resonance lies in the highly sought after terahertz to mid-infrared regime, with applications in optoelectronics\cite{freitag2013photocurrent, KoppensMueller}, optical modulators\cite{ju2011graphene, yan2012tunable, yan2013damping}, beamforming\cite{carrasco2013reflectarray}, and detection and fingerprinting of biomolecules\cite{rodrigo2015mid, hu2016far}. Enabling these applications is the ease in tuning of graphene plasmon resonances and their scattering phases through modulation of its electronic doping $n$. The modulation of $n$ can also be achieved in the temporal domain in a practical setup but its consequences on graphene plasmons are less understood. On the other hand, temporal modulation of waves has been studied in many physical contexts, revealing interesting phenomena from time-reversed acoustic\cite{Fink1}, elastic\cite{Draeger}, electromagnetic\cite{Lerosey} and water waves\cite{bacot,Przadka,Chabchoub} to the modulation of refractive index in optics\cite{Chumak,Sivan,Pendry}.

In the mid-infrared regime, the optical conductivity of graphene is well-described by the Drude model, $\sigma(\omega)=i\mathcal{D}/(\omega + i/\tau)$, where $\mathcal{D}$ is the Drude weight and $\tau$ is the electron relaxation time. In graphene, $\mathcal{D} \sim \sqrt{n}$ where $n$ is electron density, while in conventional 2D electron gas, $\mathcal{D} \sim n$. Experimental modulation of $\mathcal{D}(t)$ (or $n(t)$) can most easily be achieved with electrostatic gating\cite{fei2012gate, chen2012optical}, or via optical pumping\cite{ni2016ultrafast}. In the former, the modulation is through the change in the chemical potential, at a time scale dictated by the gate delay time in the sub-ps range. In the latter, it is via the electronic temperature, and the time scale is in the 10-100\,fs\cite{Gierz,Johannsen,LiLuo,Dawlaty}. In this letter, we examine the response of graphene plasmons under non-adiabatic temporal modulation of $\mathcal{D}(t)$, and provide the prescriptions for achieving maximal backwards propagating plasmons, standing plasmon waves, and the amplification of plasmons.

\emph{Theory} -- Consider a graphene plasmon  
\begin{equation}\label{eq:Hz}
  H_z(x, y, t) = \sign(y) e^{i\xi x -\gamma|y| - i \omega t}
\end{equation}
propagating along the $x$ axis in a sheet of graphene in the $xz$-plane. The dispersion relation for a graphene plasmon is well-known\cite{avouris20172d};
\begin{equation}\label{eq:dispersion}
  \xi = \frac{\omega}{c} \sqrt{1 - \frac{4}{\sigma(\omega)^2 Z^2}}
\end{equation}
where $c$ is the speed of light in the surrounding medium and $Z$ is the impedance of the surrounding medium. Substituting the Drude model for the conductivity and using the fact that in a typical experimental setup $Z^2 \mathcal{D}^2 \tau^2 \ll 1$ gives us
\begin{equation}
  \omega \approx -\frac{i}{2\tau} \pm \sqrt{\frac{\mathcal{D} \xi}{2\epsilon}},
\end{equation}
where $\epsilon$ is the permittivity of the surrounding medium. In general $\omega$ can be complex, where the real part corresponds to frequency and the imaginary part corresponds to damping. Note that the positive and negative square root correspond to leftward and rightward propagating plasmons, respectively.

Now we consider a time-dependent $\mathcal{D}(t)$. Under temporal modulation, $\omega$ is not a conserved quantity. On the other hand, since the graphene is spatially homogeneous, $\xi$ is invariant. Hence, one should view \eqref{eq:dispersion} as an equation for $\omega$ given $\xi$. In the quasi-static limit, $\xi \gg |\omega|$, hence, $\gamma = \sqrt{\xi^2 - \omega^2/c^2} \approx |\xi|$. Therefore, only $\omega$ in \eqref{eq:Hz} changes in time.

By discretizing $\mathcal{D}(t)$ as a series of small jumps in Drude weight we can develop a propagator matrix framework to describe the evolution of $H_z(x, y, t)$ in time. Changes in $\mathcal{D}(t)$ will cause reflection, so that
\begin{equation}
  H_z(x, y, t) = (A(t) e^{- i\omega t} + B(t) e^{i \omega^\star t}) u(x,y)
\end{equation}
where $u(x,y) = \sign(y) e^{i\xi x - \gamma |y|}$ and $\omega^\star$ is the complex conjugate of $\omega$. To understand the evolution of $H_z$ in time, we only need to keep track of the amplitude vector $[A, B]^T$ and the complex frequency $\omega$. On an interval $[t_0, t_0 + \Delta t]$, where the conductivity is constant, the amplitude vector evolves according to the propagator matrix
\begin{equation}\label{eq:D}
  M(\sigma, \Delta t) = 
  \begin{bmatrix}
    e^{-i\omega \Delta t} & 0 \\
    0 & e^{i \omega^\star \Delta t}
  \end{bmatrix}.
\end{equation}
Next, consider the change in the amplitude vector as $\mathcal{D}(t)$ undergoes a jump from $\mathcal{D}_1$ to $\mathcal{D}_2$ at $t = t_0$. In order for Maxwell's equations to be satisfied at all times we require that $H_z$ and $E_x$ be continuous at $t_0$. Using these conditions gives us the propagator matrix
\begin{equation}\label{eq:T}
  T(\mathcal{D}_1, \mathcal{D}_2) =
  \begin{bmatrix}
    \frac{1 + \omega^\star_2/\omega_1}{1+\omega^\star_2/\omega_2}
    & \frac{1 - \omega^\star_2/\omega^\star_1}{1 + \omega^\star_2/\omega_2} \\
    \frac{1 - \omega_2/\omega_1}{1 + \omega_2/\omega^\star_2}  
    & \frac{1 + \omega_2/\omega^\star_1}{1 + \omega_2/\omega^\star_2}
  \end{bmatrix}.
\end{equation}

\begin{figure}[t]
\centerline{
\includegraphics[width=0.25\textwidth]{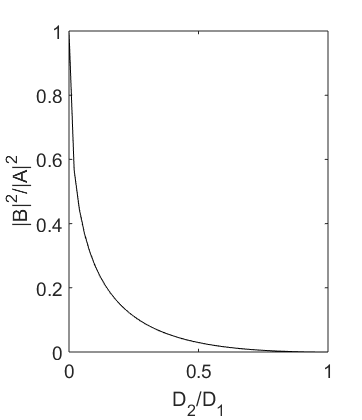}
\includegraphics[width=0.25\textwidth]{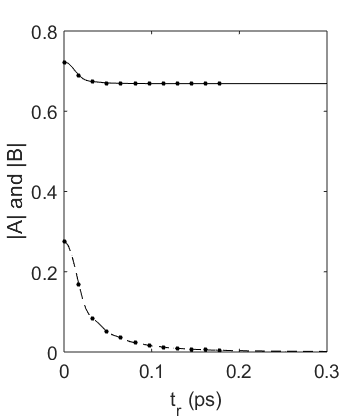}  }
\caption{(L) The ratio $|B|^2/|A|^2$ as a function of the Drude weight ratio $\mathcal{D}_2/\mathcal{D}_1$. (R) The transmission coefficients ($|A|$, solid) and reflection coefficients ($|B|$, dashes) as a function of ramp time $t_r$.  The Drude weight starts at $\mathcal{D}_1$ corresponding to Fermi level of $0.5$eV and ends at $\mathcal{D}_2$ corresponding to $0.1$eV.  Shown in dots are the values calculated by direct Maxwell simulation.}
\label{fig1}
\end{figure}

For an initially rightward propagating plasmon the amplitude vector is $[1,0]^T$. Suppose the Drude weight, initially, $\mathcal{D}_1$, goes through a jump and becomes $\mathcal{D}_2$.  If damping is small we can use the propagator to obtain the amplitudes after the jump, given by
\begin{equation}
  A = \tfrac{1}{2} \left(1 + \sqrt{\tfrac{\mathcal{D}_2}{\mathcal{D}_1}}\right),
  \quad
  B = \tfrac{1}{2} \left(1 - \sqrt{\tfrac{\mathcal{D}_2}{\mathcal{D}_1}}\right).
\end{equation}
We view $A$ and $B$ as the amplitudes of the right- and left-going waves. Thus they can be interpreted as transmission and reflection coefficients.  To maximize reflection we want $\mathcal{D}_2 \ll \mathcal{D}_1$. The ratio $|B|^2/|A|^2$ as a function of $\mathcal{D}_2/\mathcal{D}_1$ is plotted in FIG. \ref{fig1}(L). In the limit $\mathcal{D}_2/ \mathcal{D}_1\approx 0$, we achieve $|A| = |B|$, which corresponds to maximal reflection of $50\%$.

In realistic experimental setup, the Drude weight will not change instantaneously but will instead smoothly vary from $\mathcal{D}_1$ to $\mathcal{D}_2$ over some ramp time $t_r$. In this case we calculate $A$ and $B$ using both the propagator matrix method and direct full-wave simulations with excellent agreement. The graph of $|A|$ and $|B|$ for when the initial amplitude vector is $[1, 0]^T$, is shown in FIG. \ref{fig1}(R). We see that as the ramp time increases, the system moves into an ``adiabatic" regime where there is no reflection. Interestingly, the amplitude of the transmitted plasmon decreases asymptotically as it approaches the adiabatic limit. This can be understood intuitively by considering the energy density $|j|^2/(2\mathcal{D})$ of the current: a decrease in $\mathcal{D}(t)$ is countered by an increase in $|j|$. We will elaborate on this point in what follows.
\begin{figure*}[t]
\centerline{
\includegraphics[width=0.3\textwidth]{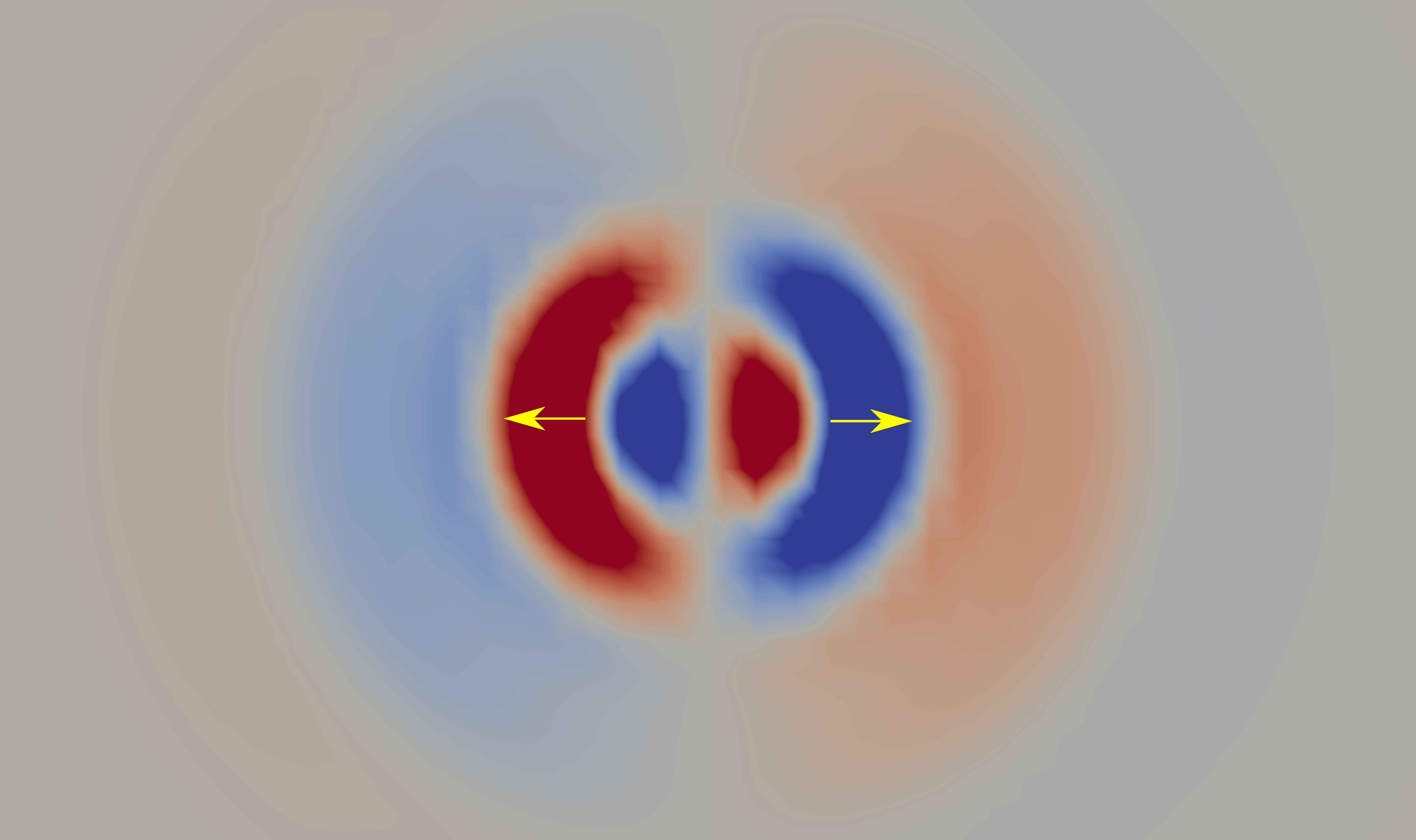}
\includegraphics[width=0.3\textwidth]{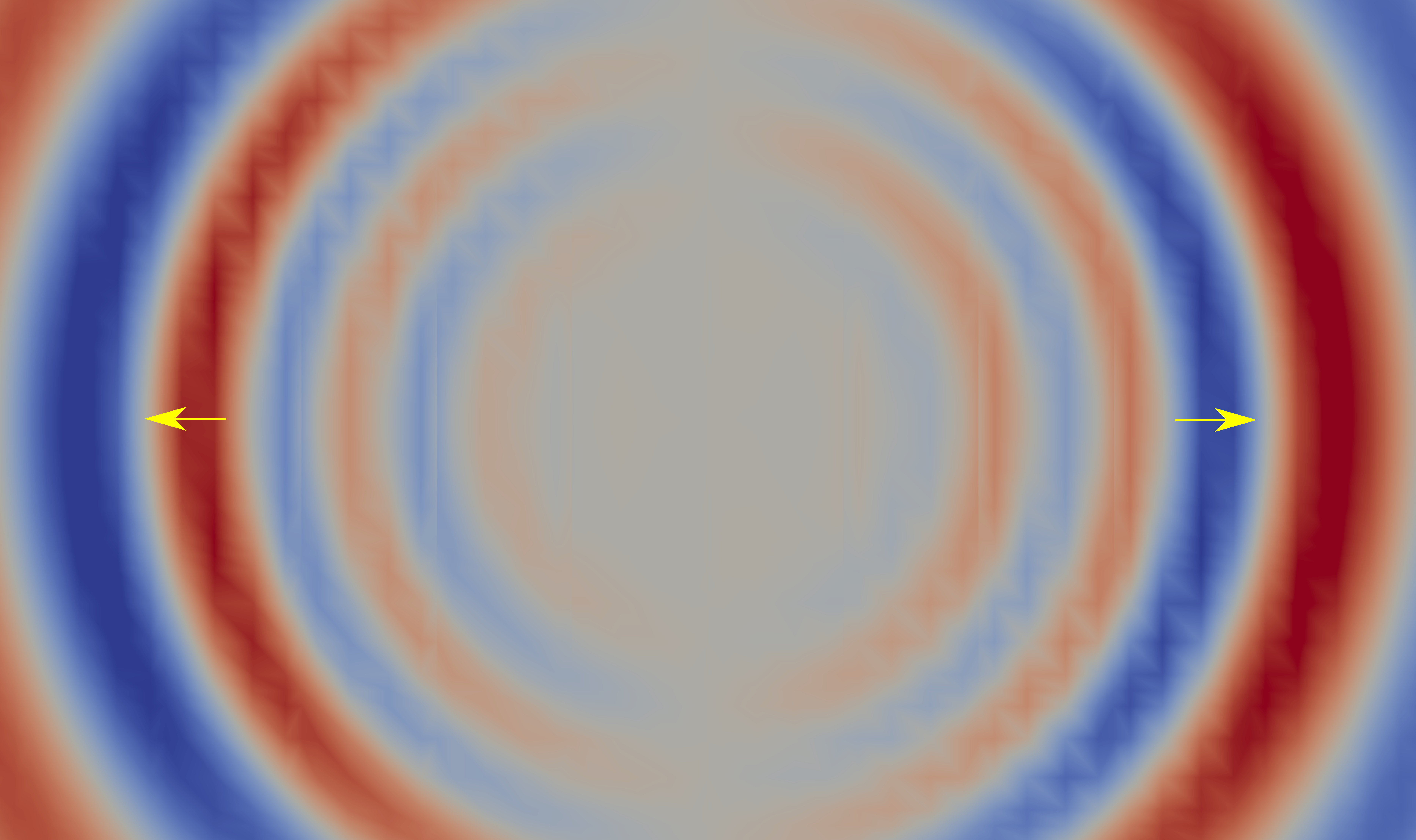}
\includegraphics[width=0.3\textwidth]{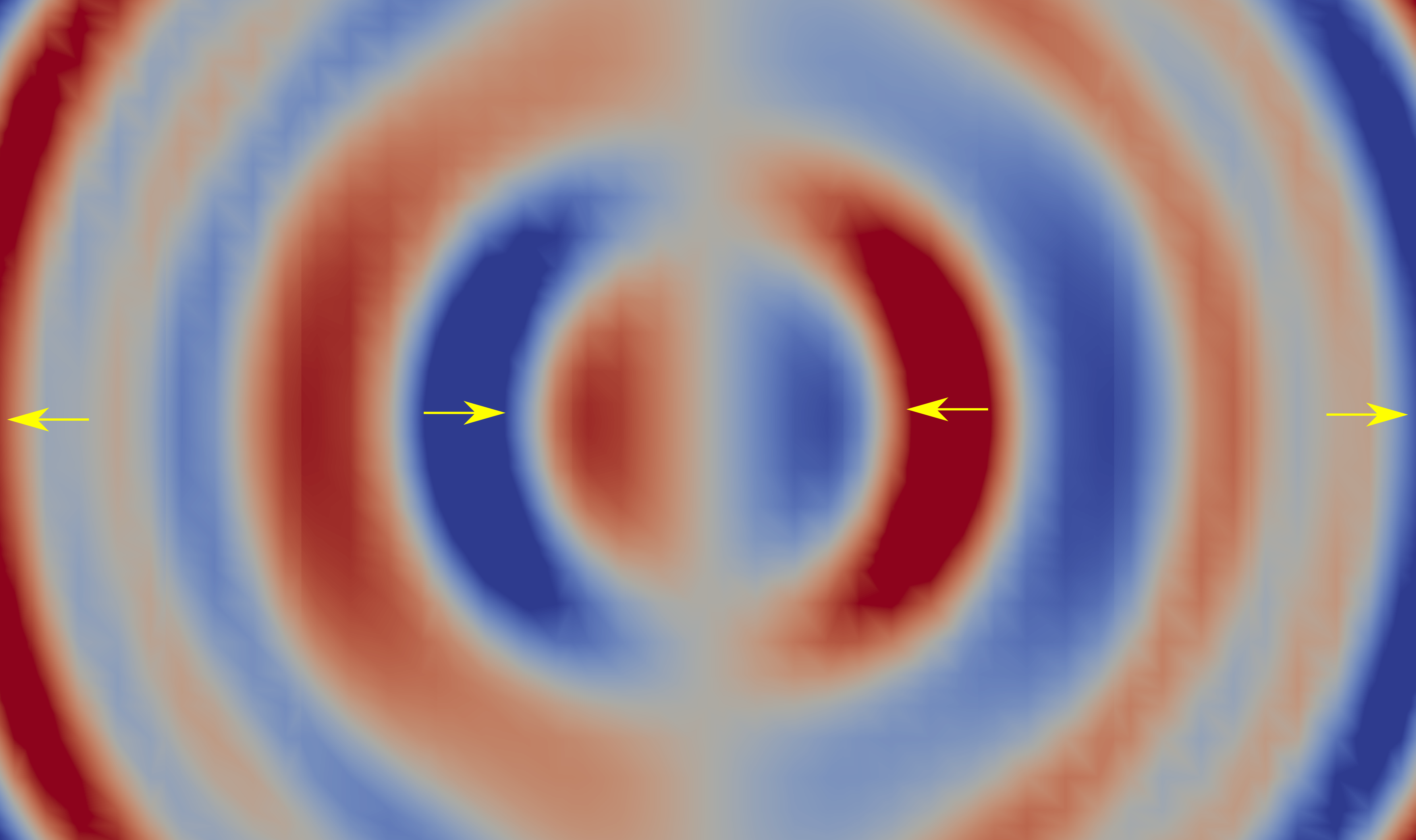}  }
\caption{Temporal reflection of graphene plasmon. Images depict the spatial distribution of $E_x$ with arrows indicating the directions of propagation. (L) A plasmon is excited with a point dipole. (C) The plasmon spreads out. (R) Right after a sudden drop in the Fermi level, part of the plasmon is reflected. In (R) the wave fronts in (C) have split into forwards and backwards propagating components with the latter effectively reverses its trajectory in time.}
\label{3dsim}
\end{figure*}

A more quantitative explanation of the decrease in transmission amplitude in the adiabatic regime can be obtained by deriving a continuum limit of the propagator matrix method. We do this in the limit where $\tau \gg 1$, in which case we have
\begin{equation}\label{eq:Tsimple}
  T(\mathcal{D}_1,\mathcal{D}_2) = \frac{1}{2}
  \begin{bmatrix}
    1 + \frac{\omega_2}{\omega_1}   
    &  1 - \frac{\omega_2}{\omega_1}  \\
    1 - \frac{\omega_2}{\omega_1}
    & 1 + \frac{\omega_2}{\omega_1}
  \end{bmatrix}
\end{equation}
where $\omega_1$ and $\omega_2$ are now real. Further note that we have $\omega_2/\omega_1 = \sqrt{\mathcal{D}_2/\mathcal{D}_1}$. In the limit where $\delta t$ is small, we can expand the propagator matrices as
\begin{align*}
  D(t + \delta t) &= 
  \begin{bmatrix}
    1 - i\omega(t) \delta t & 0 \\
    0  & 1 + i\omega(t) \delta t 
  \end{bmatrix}
  + O(\delta t^2) \\
  T(t+\delta t) &=
  \begin{bmatrix}
    1 + \frac{\mathcal{D}'(t)}{4\mathcal{D}(t)} \delta t 
    & - \frac{\mathcal{D}'(t)}{4\mathcal{D}(t)} \delta t \\
    - \frac{\mathcal{D}'(t)}{4\mathcal{D}(t)} \delta t 
    & 1 +  \frac{\mathcal{D}'(t)}{4\mathcal{D}(t)} \delta t 
  \end{bmatrix}
  + O(\delta t^2).
\end{align*}
Let $Q(t)$ be the total propagator matrix, then the change in $Q$ over a single infinitesimal conductivity step is
\begin{equation}\label{eq:Qchange}
  Q(t + \delta t) - Q(t) = [D(t + \delta t) T(t + \delta t) - I] Q(t).
\end{equation}
Dividing both sides by $\delta t$ and ignoring higher order terms, we obtain a differential equation
\begin{equation} \label{eq:Qode}
  \frac{dQ}{dt} =
  \frac{\mathcal{D}'(t)}{4\mathcal{D}(t)}
  \begin{bmatrix}
    1 & -1 \\
    -1  & 1
  \end{bmatrix}
  Q(t)
  - i\omega(t)
  \begin{bmatrix}
    1 & 0 \\
    0 & -1 
  \end{bmatrix}
  Q(t).                                                                          
\end{equation}
As $t_r \to \infty$ we obtain (see supplemental materials) the asymptotic solution
\begin{equation}
  A = e^{\frac{1}{4} \log{(\mathcal{D}(\infty)/\mathcal{D}(0))}}
  \quad
  \text{and}
  \quad
  B = 0.
\end{equation}
This limit is plotted in FIG. \ref{fig1}(R); we can see that it is in agreement with the limit obtained from the full wavelength simulations, shown in dots. Interestingly, $|A|$ can be larger than 1 if $\mathcal{D}(\infty)>\mathcal{D}(0)$. In other words, energy can be adiabatically imparted to the plasmon wave. On the other hand, when $\mathcal{D}(\infty)<\mathcal{D}(0)$, energy is being extracted instead. We will revisit these ideas later. In the adiabatic limit, $B=0$ independently of $\mathcal{D}(0)$ and $\mathcal{D}(\infty)$; hence no reflected waves. 

Experimental studies of plasmons in graphene often rely on near field optical microscopy, where plasmons are excited with an atomic scale tip\cite{fei2012gate,chen2012optical}. Hence, it is instructive to consider the propagation of a 2-D plasmon wave excited by a point dipole under temporal modulation of its Drude weight. We performed a full 3-D Maxwell simulation of this setup using NekCEM\cite{nekCEM}. Here, plasmons wave emitted by the point dipole propagates radially outwards, which upon an instantaneous change in the Fermi level, results in a reflected wave that propagates inwards and refocuses back to its point of origin. Effectively, we have a `time mirror', which reflects the wave back in time, much akin to a spatial discontinuity that reflects the wave.  The resulting spatial distribution of $E_x$ at different times are depicted FIG. \ref{3dsim}. 

The concept of `time mirror' has been discussed in various context of waves phenomena\cite{Fink1,Draeger,Lerosey,Przadka,Chabchoub}.
Recently, the `time mirror' has been observed in the context of water waves, showing clear reversal of shallow water waves\cite{bacot}.  In this experiment, a circular wave is generated with a point source.  While the wave expands, the tank is accelerated in an almost instantaneous fashion.  The acceleration interacts with the expanding wave and generates a transmitted component and reflected component.  The former is a circular wave whose radius continues to expand, while the latter is a circular wave with decreasing radius.

\emph{Plasmon amplification} -- We have already observed that by ramping up the Drude weight from $\mathcal{D}(0)$ to $\mathcal{D}(\infty)$ we can create a transmitted wave whose coefficient is greater than 1.  We wish to explore how $\mathcal{D}(t)$ can put energy into the system.  By applying Stoke's identity to the two-dimensional Maxwell system, we arrive at the identity
\begin{align*}
\frac{d}{dt} &\left[ 
      \int \frac{\epsilon}{2}\left(|E_x|^2 + |E_y|^2\right) +
                        \frac{\mu}{2}|H_z|^2 \; dxdy  
            +  \int \frac{1}{2\mathcal{D}} |j|^2 dx 
             \right] \\
&= -\left[ \frac{\mathcal{D}'}{\mathcal{D}} + \frac{2}{\tau} \right]
\int \frac{1}{2\mathcal{D}} |j|^2 dx,
\end{align*}
where the $x$ integration is over one period of the plasmon and the $y$ integration is from $-\infty$ to $\infty$. We see that it is possible to inject energy into the system by modulating $\mathcal{D}(t)$ and producing a positive right-hand side. 
  
\begin{figure}[b]
\includegraphics[width=0.4\textwidth]{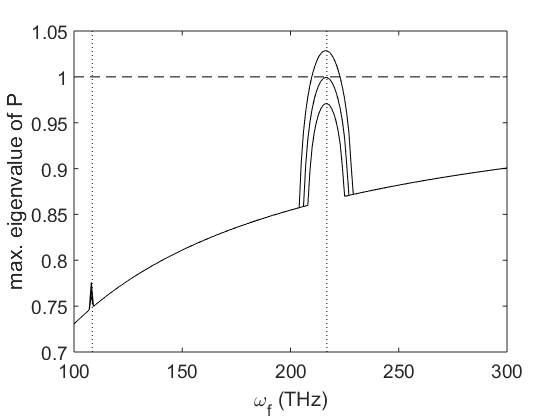}
\caption{The maximum eigenvalue of the propagator matrix $P$ after one period of the sinusoidal Drude weight $\mathcal{D}(t)=\mathcal{D}_0 + \Delta\mathcal{D}\sin \omega_f t$ as a function of forcing frequency $\omega_f$.  Here $\mathcal{D}_0$ corresponds to Fermi level of $0.5$eV and $\Delta\mathcal{D}$ is chosen so that it is $80\%$, $100\%$, and $120\%$ of the critical value.  When the maximum eigenvalue is greater than one, there is amplification.  Observe also that the maximum amplification occurs at $\omega_f= 2\omega_0$; both $\omega_0$ and $\omega_f$ are indicated with vertical dotted lines.} 
\label{fig:parametric-amp}
\end{figure}

The increase in energy can be understood in terms of parametric resonance\cite{landaulifshitz}.  We set this up by considering the electron density on the plasmon on the graphene after removing the oscillatory $x$-dependence.  The electron density (amplitude) satisfies
\[
\frac{d^2 n}{dt^2} + \frac{1}{\tau} \frac{dn}{dt} + \frac{\xi} {2\epsilon} \mathcal{D}(t) n = 0,
\]
in the quasi-static limit.  We excite the system by modulating the Drude weight as
\[
\mathcal{D}(t) = \mathcal{D}_0 + \Delta \mathcal{D} \sin (\omega_f t).
\]
From parametric resonance theory, we predict that growth is expected when $\omega_f = 2\omega_0$, where $\omega_0 = \sqrt{D_0\xi/(2\epsilon)}$ is the plasmon frequency for large $\tau$.  Amplification overcomes damping when
\[
\frac{\Delta \mathcal{D}}{\mathcal{D}_0} > \frac{2}{\tau\omega_0}.
\]

\begin{figure*}[t]
\centerline{
\includegraphics[width=0.4\textwidth]{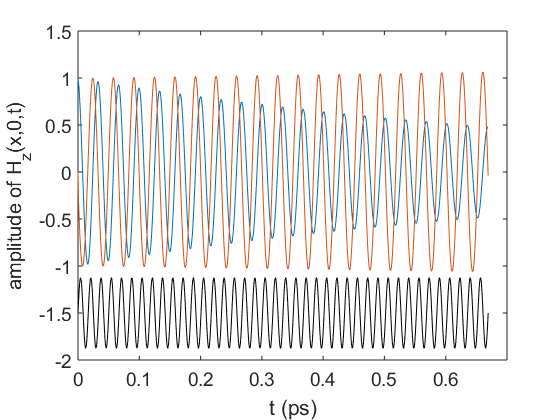}
\includegraphics[width=0.4\textwidth]{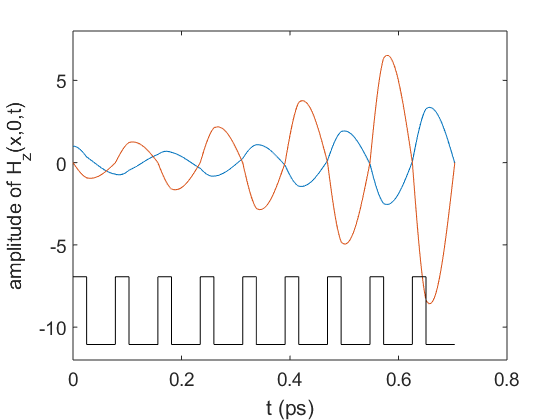}  }
\caption{The graphs of the real (blue) and imaginary (red) parts of the amplitude of $H_z(x,0,t)$, namely $A(t) e^{-i\omega t} + B(t) e^{i\omega^\star t}$. (L) Under sinusoidal $\mathcal{D}(t)$ where the Drude weight has a mean corresponding to Fermi level of $0.5$eV.  Its sinusoidal amplitude is at 120\% of the critical value. The growth of the imaginary part is visible.  Shown at the bottom (not in scale) is the periodic excitation at frequency $2\omega_0$. (R) Under periodic piecewise constant $\mathcal{D}(t)$ where the Drude weight alternates between Fermi energies of $E_1=0.1$eV and $E_2=0.05$eV, and over time intervals $t_1=25$fs and $t_2=53$fs.  In this calculation, $\epsilon=3$.  Observe that the real and imaginary parts of the expression are in antiphase, thus corresponding to a standing wave.}
\label{fig3}
\end{figure*}

To verify the theory, we consider a sinusoidal Drude weight time dependence and computed the propagator matrix $P$ for a single period.  The maximum eigenvalues of the matrix determines if amplification takes place.  Amplification occurs when the maximum eigenvalue is greater than 1.  For the verification, we plot the maximum eigenvalue of the propagator as a function of forcing frequency $\omega_f$ for values of $\Delta \mathcal{D}$ below, equal to, and above the critical value for amplification.  The resulting graphs, shown in FIG. \ref{fig:parametric-amp}, confirm our prediction. 

To further understand the phenomenon of amplification, we consider exciting the plasmon sinusoidally at frequency $\omega_f=2\omega_0$ and $\Delta \mathcal{D}/\mathcal{D}_0$ corresponding to $120$\% above the critical value for amplification.  The plasmon is initially right-going so its amplitude vector is $[1,0]^T$.  We graph the real and imaginary parts of the amplitude of $H_z(x,0,t)$, i.e., $A(t)e^{-i\omega t} + B(t)e^{i\omega^\star t}$, as a function of time.  We observe that there is a noticeable growth in the imaginary part of the amplitude, whereas the real part appears to be decreasing.  Upon closer inspection, the modulus of the amplitude does exceed 1 whenever the imaginary part hits its peak or trough.

We next investigate whether we can produce richer control of amplification by altering the time dependence of $\mathcal{D}(t)$.  We consider the simple case where $\mathcal{D}(t)$ is piecewise constant and periodic. The period is of the Drude weight is $t_1 + t_2$, wherein within a period 
\[
\mathcal{D}(t) = \left\{ \begin{array}{ll}
                        \mathcal{D}_1  & \mbox{for} \;\; 0 < t < t_1 , \\
                        \mathcal{D}_2   & \mbox{for}  \;\; t_1 < t < t_1+t_2 .
                           \end{array}
                           \right.
\]
For this part, we keep the damping finite, leading to complex frequencies. 
In keeping with our previous notation, the propagator from $t=0$ to $t=t_1$ is
\[
M_1 = \left[ \begin{array}{cc}
    e^{-i\omega_1 t_1} & 0 \\
   0  & e^{i \omega^\star_1 t_1} 
                         \end{array}
                \right] .
\]
Across the interface at $t=t_1$, the propagator is
\[
T_1 =  \left[ \begin{array}{cc}
\frac{1 + \omega^\star_2/\omega_1}{1+\omega^\star_2/\omega_2}   
        &  \frac{1 - \omega^\star_2/\omega^\star_1}{1+\omega^\star_2/\omega_2}   \\
\frac{1 - \omega_2/\omega_1}{1+\omega_2/\omega^\star_2}  
       &  \frac{1 + \omega_2/\omega^\star_1}{1+\omega_2/\omega^\star_2}
                                           \end{array}
                                  \right] .
\]
Similarly, we have $M_2$ and $T_2$ corresponding to propagation from $t_1$ to $(t_1+t_2)$ and across the interface at $(t_1+t_2)$.
The propagator for a single period is $P = T_2 M_2 T_1 M_1$.
If $\mathcal{D}(t)$ goes through $N$ periods, the propagator is ${\bf P} = P^N$.  We note that the matrix ${\bf P}$ depends on $t_1$, $t_2$, $\mathcal{D}_1$ and $\mathcal{D}_2$.  We expect it to exhibit different behavior depending on these parameters.

To analyze the properties of ${\bf P}$, we diagonalize $P$ and write $P = V \Lambda V^{-1}$, so that ${\bf P} = V \Lambda^N V^{-1}$.  We start with a right-going wave, i.e. initial vector is $[1, 0]^{T}$.  Let $\lambda_1 $ and $\lambda_2$ be the eigenvalues of $P$ and set $W = V^{-1}$.  Then the right- and left-going components after $N$ periods can be found by examining the first column of ${\bf P}$
\[
\left[ \begin{array}{c}
         {\bf P}_{11} \\ {\bf P}_{21} 
         \end{array} \right] = V \Lambda^N W 
\left[ \begin{array}{c}
         1 \\ 0 
         \end{array} \right] 
\]
Multiplying out, we have
\[
\left[ \begin{array}{c}
         {\bf P}_{11} \\ {\bf P}_{21} 
         \end{array} \right] =
\left[ \begin{array}{c}
         \lambda_1^N V_{11} W_{11} + \lambda_2^N V_{12} W_{21} \\ 
         \lambda_1^N V_{21} W_{11} + \lambda_2^N V_{22} W_{22} 
         \end{array} \right] .
\]

Denote the entries of $P$ by
\[
P = \left[ \begin{array}{cc}
              c & d \\
            \overline{d} & \overline{c} 
               \end{array} \right] .
\]
The eigenvalues of $P$ are
\newcommand{\imag}{\mbox{Im\,}}
\newcommand{\real}{\mbox{Re\,}}
\[
\lambda_{1,2} = \real c \pm \sqrt{ |d|^2 - ( \imag c )^2 } .
\]
If $(\imag c )^2 \geq |d|^2$ then $|\lambda_1| = |\lambda_2|$.  Otherwise both eigenvalues are real and $\lambda_1 > \lambda_2$.  We further assume that $\lambda_1 > 1$ and $\lambda_2 < 1$ for some
$\mathcal{D}_1$, $\mathcal{D}_2$, $h_1$ and $h_2$.  For $N$ large, the terms involving $\lambda_2$ can be dropped, giving
\[
\left[ \begin{array}{c}
         {\bf P}_{11} \\ {\bf P}_{21} 
         \end{array} \right] \approx \lambda_1^N
\left[ \begin{array}{c}
         V_{11} W_{11}  \\ 
         V_{21} W_{11}  
         \end{array} \right] .
\]
Here, ${\bf P}_{11}$ corresponds to right-going (transmitted) wave and ${\bf P}_{21}$, the left-going (reflected) wave.  We can make these components as large as we like as long as $\lambda_1>1$.

We can solve for the column of $Q$ corresponding to $\lambda_1$ from $(P - \lambda_1) Q = I$.  We get
\[
V_{11} = d, \;\;\; V_{21} = -i \imag c + \sqrt{|d|^2 - (\imag c)^2 } ,
\]
so that $|V_{21}| = |d|^2$ if $|d|^2 - (\imag c)^2 > 0$.  This has interesting implications.

To make a field consisting of large transmission and small reflection that is continuously amplified as $\mathcal{D}$ is cycled, we must find parameters such that
\[
\frac{|V_{11}|}{|V_{21}|} \; \mbox{large, and} \;\; \lambda_1 >1 .
\]
However, we see that this is impossible as this ratio will be fixed at 1 when $\lambda_1>1$.  Similarly, we cannot make a field consisting of large reflection and small transmission that is amplified.

However, it is possible to make a field that consists mostly of a standing wave that grows.  For this to happen, we must find parameters such that
\[
| V_{11} - V_{21} | \;\; \mbox{or} \;\; | V_{11} + V_{21} | \;\; \mbox{small, and} \;\; \lambda_1 > 1 .
\] 
We were able to find parameter settings where this is true.  Thus it is possible to generate growing standing waves as shown in FIG. \ref{fig3}(R).

\emph{\label{sec:level5}Conclusion} -- In summary, we discussed how time modulation of the plasmonic Drude weight can enable rich control of plasmons in space and time, such as inducing reversed trajectory backwards in time, producing standing waves, and overcoming loss to achieve amplification. Our estimates, considering experimentally feasible parameters, suggest that these phenomena should be observable.

\emph{\label{sec:level6}Acknowledgement} -- This work was initiated at the Institute for Mathematics and its Applications (IMA).  JW and MM acknowledge funding from U.S. Department of Energy, Office of Science, under contract DE-AC02-06CH11357. FS acknowledges support from NSF awards DMS-DMS-1211884 and DMS-1440471. TL acknowledges support from IMA and NSF/EFRI- 1741660.

\newpage

\newpage
\onecolumngrid
\appendix
\section{Deriving asymptotes of scattering coefficient}
The asymptotes in FIG. \ref{fig1}(R) can be explained by deriving a continuum limit of the propagator matrix method. We start with the `propagator matrix' taking right-going and left-going waves across an interface and propagating it a distance of $\delta t$.  The interface is at time $t_i$.  To the left of the interface, the Drude weigtht is $\mathcal{D}_i$; to the right, it is $\mathcal{D}_{i+1}$.  If damping is negligible, the complex frequencies have zero imaginary parts.  Let us denote the frequency to the left of the interface by $\omega_i$ and to the right, by $\omega_{i+1}$.  The matrix corresponding to crossing the interface [see \eqref{eq:Tsimple}] is given by
\[
T_{i+1} = \frac{1}{2} \left[ \begin{array}{cc}
1 + \frac{\omega_{i+1}}{\omega_i}   &  1 - \frac{\omega_{i+1}}{\omega_i}  \\
1 - \frac{\omega_{i+1}}{\omega_i}   &  1 + \frac{\omega_{i+1}}{\omega_i} 
                                           \end{array}
                                  \right] .
\]
To propagate the state a distance of $\delta t$ in the medium where frequency is $\Omega_{i+1}$, the matrix required [see \eqref{eq:D}] is
\[
M_{i+1} = \left[ \begin{array}{cc}
    e^{-i\omega_{i+1} \delta t} & 0 \\
   0  & e^{i\omega_{i+1} \delta t} 
                         \end{array}
                \right] .
\]
The frequencies and conductivities satisfy
\[
\omega_{i+1} = \sqrt{\frac{\mathcal{D}{i+1}}{\mathcal{D}_{i}}  }  \omega_i.
\]
Let the combined propagator be denoted by $P_{i+1}$, given by
\[
P_{i+1} = M_{i+1} T_{i+1} .
\]

Suppose for $t < 0$, $\mathcal{D} = \mathcal{D}_0$.  At $t = 0, \delta t, 2\delta t, \cdots, N\delta t$, $\mathcal{D}$ transitions abruptly to $\mathcal{D}_1, \mathcal{D}_2, \cdots, \mathcal{D}_N$.  Then the propagator for the stack of $N$ `slabs' is
\[
Q_N = P_{N} P_{N-1} P_{N-2} \cdots P_2 P_1 .
\]
Instead of a set of matrix products, we consider iterations of the form
\[
Q_{i+1} = P_{i+1} Q_i, 
\]
with $Q_1 = P_1$.  Letting $t_0 = 0$, after $N$ slabs, $t_N = N \delta t =:T$.  We are interested in the behavior of $Q_N$ as both $N$ and $T$ go to infinity while $\mathcal{D}$ remain bounded.  With this in mind, we rewrite the above as
\[
Q_{i+1} - Q_i = (P_{i+1} - I ) Q_i .
\]
Associated with index $i$ is time $t=i\delta t$.  Therefore index $i+1$ is associated with time $t+\delta t$.  We relabel $T_{i+1}$ as $T(t + \delta t)$ and write
\[
T(t+\delta t) = \frac{1}{2} \left[ \begin{array}{cc}
                               1 + \frac{\omega(t+\delta t)}{\omega(t)} &  1 - \frac{\omega(t+\delta t)}{\omega(t)} \\
                               1  - \frac{\omega(t+\delta t)}{\omega(t)} &  1 + \frac{\omega(t+\delta t)}{\omega(t)}
                                          \end{array}
                                         \right] ,
\]
with the understanding that $\omega(t) = \omega_i$ and $\omega(t+\delta t) = \omega_{i+1}$. Similarly, we write
\[
D(t+\delta t) = \left[ \begin{array}{cc}
    e^{-i\omega(t+\delta t) \delta t} & 0 \\
   0  & e^{i\omega(t+\delta t) \delta t} 
                         \end{array}
                \right] .
\]
For small $\delta t$, we have
\begin{align*}
\omega(t+\delta t) \delta t & = [ \omega(t) + \omega'(t) \delta t + O(\delta t^2 ) ] \delta t \\
                                          & = \omega(t) \delta t + O(\delta t^2)  .
\end{align*}
Therefore, we can expand $D(t+\delta t)$ as
\[
D(t+\delta t) =  \left[ \begin{array}{cc}
    1 - i\omega(t) \delta t & 0 \\
   0  & 1 + i\omega(t) \delta t 
                         \end{array}
                \right]  + O(\delta t^2).
\]
Similarly, we expand $T(t+\delta t)$ for small $\delta t$ by first observing that
\begin{align*}
\frac{\omega(t+\delta t)}{\omega(t)} &= \sqrt{ \frac{\mathcal{D}(t+\delta t)}{\mathcal{D}(t)}  }\\
                                                           & = \sqrt{\frac{\mathcal{D}(t) + \mathcal{D}'(t) \delta t + O(\delta t^2)}{\mathcal{D}(t)} }\\
                                                         & = 1 + \frac{\mathcal{D}'(t)}{2\mathcal{D}(t)} + O(\delta t^2) .
\end{align*}
Therefore, we can write $T(t+\delta t)$ as
\[
T(t+\delta t) = \left[ \begin{array}{cc}
                  1 + \frac{\mathcal{D}'(t)}{4\mathcal{D}(t)} \delta t  &  - \frac{\mathcal{D}'(t)}{4\mathcal{D}(t)} \delta t \\
                      - \frac{\mathcal{D}'(t)}{4\mathcal{D}(t)} \delta t  & 1 +  \frac{\mathcal{D}'(t)}{4\mathcal{D}(t)} \delta t 
                                \end{array}
                        \right]  + O(\delta t^2) .
\]
Multiplying we get $P(t+\delta t)$ as
\begin{align*}
&P(t+\delta t) = \\
&\left[ \begin{array}{cc}
          1 + \left(\frac{\mathcal{D}'(t)}{4\mathcal{D}(t)} - i\omega(t) \right) \delta t & -\frac{\mathcal{D}'(t)}{4\mathcal{D}(t)} \delta t \\
          -\frac{\mathcal{D}'(t)}{4\mathcal{D}(t)} \delta t & 1 + \left(\frac{\mathcal{D}'(t)}{4\mathcal{D}(t)} + i\omega(t) \right) \delta t 
                                \end{array}
                         \right] + O(\delta t^2) .
\end{align*}
The iteration for $Q_i$ is now rewritten as
\[
Q(t+\delta t) - Q(t) = (P(t+\delta t) - I) Q(t).
\]
Dividing both sides by $\delta t$ and ignoring $O(\delta t)$, we obtain an ODE for $Q(t)$
\begin{equation} \label{eq:Qeqn}
\frac{d Q}{dt} = \frac{\mathcal{D}'(t)}{4\mathcal{D}(t)} \left[ \begin{array}{rr}
                                                                                      1  & -1 \\
                                                                                     -1  &  1 
                                                                                \end{array} \right]  Q(t)
                               - i\omega(t) \left[ \begin{array}{rr}
                                                      1 & 0 \\
                                                      0 & -1 
                                                            \end{array} \right] Q(t) .                                                                          
\end{equation}

To solve \eqref{eq:Qeqn} start by rescaling time to $t = s t_r$ where $t_r$ is the total ramp time transitioning the Drude weight from $\mathcal{D}(0)$ to $\mathcal{D}(t_r)$.  Then (\ref{eq:Qeqn}) can be rewritten as
\begin{equation}\label{eq:Qeqnscaled}
\frac{dQ}{ds} = t_r \left\{ \left[ \begin{array}{cc}
                                     - i \omega & 0 \\
                                       0        & i\omega 
                                              \end{array} \right]  + \frac{\mathcal{D}'(s)}{4 t_r \mathcal{D}(s)} \left[ \begin{array}{rr}
                                                                                      1  & -1 \\
                                                                                     -1  &  1 
                                                                                \end{array} \right] \right\} Q .
\end{equation}
Following Coddington and Levinson (Chapter 7) \cite{cod-lev}, we see a solution matrix of the form
\[
Q(s) = P(s) e^\Phi ,
\]
where
\[
P = P_0 + \frac{1}{t_r} P_1 + \frac{1}{t_r^2} P_2 + \cdots ,
\]
and
\[
\Phi = t_r \Phi_0 + \Phi_1,
\]
where $\Phi$ is diagonal. Now we can substitute the solution into (\ref{eq:Qeqnscaled}) and collect terms of the same order in $t_r$.

The $O(T)$ equation is straight-forward to solve since $A_0$ is diagonal.  The solution is
\[
\Phi'_0 = \left[ \begin{array}{cc}
                                     - i \omega & 0 \\
                                            0        & i\omega 
                                              \end{array} \right] ,
\]
and $P_0 = I$.  Therefore, $P'_0 = 0$.  In addition, we have
\[
\Phi'_1 + P_1 \left[ \begin{array}{cc}
                                     - i \omega & 0 \\
                                            0        & i\omega 
                    \end{array} \right] 
      = \left[ \begin{array}{cc}
                                     - i \omega & 0 \\
                                            0        & i\omega 
                    \end{array} \right]  P_1 + \frac{\mathcal{D}'(s)}{4 \mathcal{D}(s)} \left[ \begin{array}{rr}
                                                                                      1  & -1 \\
                                                                                     -1  &  1 
                                                                                \end{array} \right] .
\]
Looking only at the diagonal terms of the above equation, we see that
\[
\Phi'_1 = \frac{\mathcal{D}'(s)}{4 \mathcal{D}(s)} \left[ \begin{array}{rr}
                                                                                      1  &  0 \\
                                                                                      0  &  1 
                                                                                \end{array} \right] .
\]
Since we are only interested in the solution for large $t_r$, we need not go further in solving for $P_1$.  The solution matrix is
\begin{align*}
Q(s) =& \left(P_0 + \frac{1}{t_r} P_1 \right) \exp ( t_r \Phi_0 + \Phi_1 ) \\
       = & \left( I + \frac{1}{t_r} P_1 \right) \left[ \begin{array}{cc} 
                                                        e^{-i t_r \int_0^s \omega(s') ds'} & 0 \\
                                                        0 & e^{i t_r \int_0^s \omega(s') ds'} 
                                                        \end{array} \right]  \\
          & \hspace{4em}            \left[ \begin{array}{cc}
                                                       e^{ \frac{1}{4} (\log \mathcal{D}(s) - \log \mathcal{D}(0) ) } & 0 \\
                                                       0 & e^{ \frac{1}{4} (\log \mathcal{D}(s) - \log \mathcal{D}(0) ) }
                                                        \end{array} \right] .
\end{align*} 
 Consider the first column of $Q(t_r)$, which is given by
\[
Q_1(T) = \left[ \begin{array}{c}
                      \left(1 + \frac{P_{1,11} }{t_r} \right) e^{-i t_r \int_0^{t_r} \omega(s') ds'} 
                                                e^{ \frac{1}{4} \log{\frac{\mathcal{D}(t_r)}{\mathcal{D}(0)} }  }\\
                       0 
                        \end{array} \right] .
\]        
The entries correspond to transmitted and reflected waves when the incident wave is a right-going wave of unit amplitude.  Therefore, the transmission coefficient is the modulus of the first entry, i.e.,
\[
A = \left| 1 + \frac{P_{1,11} }{t_r} \right|    e^{ \frac{1}{4} \log{\frac{\mathcal{D}(t_r)}{\mathcal{D}(0)} } }  ,
\]
which simplifies to 
\[
A = e^{ \frac{1}{4} \log{\frac{\mathcal{D}(\infty)}{\mathcal{D}(0)} }  } ,
\]
when $t_r \rightarrow \infty$.  This explains the asymptotic limits of the transmission coefficient and the reflection coefficient (zero) as the ramp time becomes large.

\end{document}